\documentclass[pra,twocolumn,superscriptaddress]{revtex4}
\usepackage{amsfonts}
\usepackage{amssymb}
\usepackage{amsmath}
\usepackage{epsfig}
\usepackage{color}
\usepackage{graphics, graphicx}
\usepackage{bbold}
\usepackage{psfrag}
\usepackage{mathcomp}
\usepackage{subfigure}
\usepackage{verbatim}
\usepackage[colorlinks,citecolor=blue]{hyperref}

\makeatletter

\newcommand{\Rmnum}[1]{\expandafter\@slowromancap\romannumeral #1@}
\makeatother

\begin{document}

\title{Topological optical Raman superlattices}

\author{Jia-Hui Zhang}
\affiliation{State Key Laboratory of Quantum Optics and Quantum Optics Devices, Institute
of Laser Spectroscopy, Shanxi University, Taiyuan, Shanxi 030006, China}
\affiliation{Collaborative Innovation Center of Extreme Optics, Shanxi
University,Taiyuan, Shanxi 030006, China}

\author{Bei-Bei Wang}
\affiliation{State Key Laboratory of Quantum Optics and Quantum Optics Devices, Institute
of Laser Spectroscopy, Shanxi University, Taiyuan, Shanxi 030006, China}
\affiliation{Collaborative Innovation Center of Extreme Optics, Shanxi
University,Taiyuan, Shanxi 030006, China}

\author{Feng Mei}
\email{meifeng@sxu.edu.cn}
\affiliation{State Key Laboratory of Quantum Optics and Quantum Optics Devices, Institute
of Laser Spectroscopy, Shanxi University, Taiyuan, Shanxi 030006, China}
\affiliation{Collaborative Innovation Center of Extreme Optics, Shanxi
University,Taiyuan, Shanxi 030006, China}

\author{Jie Ma}
\affiliation{State Key Laboratory of Quantum Optics and Quantum Optics Devices, Institute
of Laser Spectroscopy, Shanxi University, Taiyuan, Shanxi 030006, China}
\affiliation{Collaborative Innovation Center of Extreme Optics, Shanxi
University,Taiyuan, Shanxi 030006, China}

\author{Liantuan Xiao}
\affiliation{State Key Laboratory of Quantum Optics and Quantum Optics Devices, Institute
of Laser Spectroscopy, Shanxi University, Taiyuan, Shanxi 030006, China}
\affiliation{Collaborative Innovation Center of Extreme Optics, Shanxi
University,Taiyuan, Shanxi 030006, China}

\author{Suotang Jia}
\affiliation{State Key Laboratory of Quantum Optics and Quantum Optics Devices, Institute
of Laser Spectroscopy, Shanxi University, Taiyuan, Shanxi 030006, China}
\affiliation{Collaborative Innovation Center of Extreme Optics, Shanxi
University,Taiyuan, Shanxi 030006, China}

\date{\today }

\begin{abstract}
  Topological phases of ultracold atoms recently have been intensively studied both in optical superlattices and Raman lattices. However, the topological features induced by the interplay between such two lattices remain largely unexplored. Here we present an optical Raman superlattice system that incorporates an optical superlattice and a Raman superlattice. The Raman superlattice presented here supports tunable dimerized spin-orbit couplings and staggered on-site spin flips. We find that such system respects a spin-rotation symmetry and has much richer topological properties. Specifically, we show that various topological phases could emerge in the optical Raman superlattice, such as four different chiral topological insulator phases and two different quantum spin Hall insulator phases, identified by spin winding and spin Chern numbers respectively. We also demonstrate that the spin-dependent topological invariants could be directly measured by quench dynamics.

\end{abstract}

\maketitle

\section{Introduction}

Ultracold atoms trapped in optical lattices provide a versatile synthetic system for exploring topological phases of matter~\cite{TPCArev1,TPCArev2,TPCArev3}. Many pioneering experimental progresses achieved in this field are simply based on optical superlattices. For instance, using ultracold atoms trapped in one-dimensional (1D) optical superlattices,
the seminal Su-Schrieffer-Heeger (SSH) model~\cite{SSH} has been naturally implemented in experiment~\cite{SSHexp}.
It is well known that such model supports two-band topological insulator phases protected by chiral symmetry~\cite{SSH}. The quantized Zak phases characterizing the topological features of the Bloch bands have been directly measured by Bloch oscillations and Ramsey interferometry~\cite{SSHexp}. Furthermore, quantized topological pumping~\cite{Thouless} also has been experimentally realized~\cite{Bloch2016,Takahashi2016,Spielman2016,Bloch2018} by controlling the optical superlattices in a cyclic and adiabatic manner~\cite{Zhang2011,Wang2013,Mei2014,Wei2015,Lee2017,Hayward2018,Mei2019,Lee2019,Lee2020,Ke2020,Chen2020}. In addition to 1D topological phenomena, optical superlattice system also allows to explore 2D~\cite{Lang2012,Mei2012} and 4D integer quantum Hall insulator phases~\cite{Bloch2018}.

In parallel, with the experimentally successful preparation of spin-orbit couplings in ultracold gases~\cite{SOC1,SOC2,SOC3,SOC4}, optical Raman lattices have been developed into a powerful platform for implementing spin-orbit couplings in lattice systems~\cite{Liu2013,Liu2014,Pan2016,Liu2018,Pan2018,Song2018,CI2018,CI2019,Song2019,Pan2021}, in which Raman lattice potentials are additionally applied except the conventional optical lattice trapping potentials. The Raman potentials are generated through a two-photon Raman transition which couples the spin up and spin down encoded by two atomic internal states~\cite{Liu2013,Liu2014}. Based on generated spin-orbit couplings in different dimensions, various topological phases could be created and probed in optical Raman lattices, including 1D topological insulator phases~\cite{Song2018}, 2D topological Chern insulator phases~\cite{Pan2016,Pan2018,CI2018,CI2019} and 3D topological Weyl~\cite{Pan2021,3DWeyl1,3DWeyl2,3DWeyl3} and nodal-line semimetal phases~\cite{Song2019,3DNL1,3DNL2}. Moreover, our recent study also shows that optical Raman lattices have the ability to generate complex spin-orbit couplings by designing suitable Raman lattice potentials~\cite{Mei2021}, like the 3D next-nearest-neighbor spin-orbit couplings, that could enable exotic topological nodal chain semimetal phases~\cite{Mei2021}.

In this paper, we present an optical Raman superlattice system that integrates an optical superlattice and a Raman superlattice. Different from previous Raman lattices, the Raman superlattice could generate dimerized spin-orbit couplings and staggered on-site spin flips, that can be directly implemented through two proper Raman lasers, which has not been reported before. We find that the interplay between the optical superlattice and the Raman superlattice could lead to much richer topological phases. First, we reveal that the optical Raman superlattice system satisfies a spin-rotation symmetry, that allows us to use spin-dependent topological invariants to identify its topological property. Second, we demonstrate that the system in the case of turning off the on-site terms could support four different 1D four-band chiral topological insulator phases, characterized by spin winding numbers. Third, with on-site terms, the corresponding system could be mapped into a synthetic 2D momentum space and supports two different quantum spin Hall insulator phases and one double Chern insulator phase, characterized by spin Chern numbers. In both cases, the topological properties are detailedly explored by numerically extracting topological phase diagrams and demonstrating bulk-edge correspondences. In addition, we also show that both the spin winding and spin Chern numbers could be directly measured by quench dynamics.

The paper is organized as follows. Section II presents the construction of optical Raman superlattices. Section III studies the symmetry of the Bloch Hamiltonian. Sections IV and V exhibits that optical Raman superlattices provide a versatile platform for exploring various four-band 1D and 2D topological insulator phases, protected by spin winding and spin Chern numbers respectively. Section VI summaries the main results of this paper and outlines future works along this line.

\begin{figure}
	\centering
	\includegraphics[width=8cm,height=4.1cm]{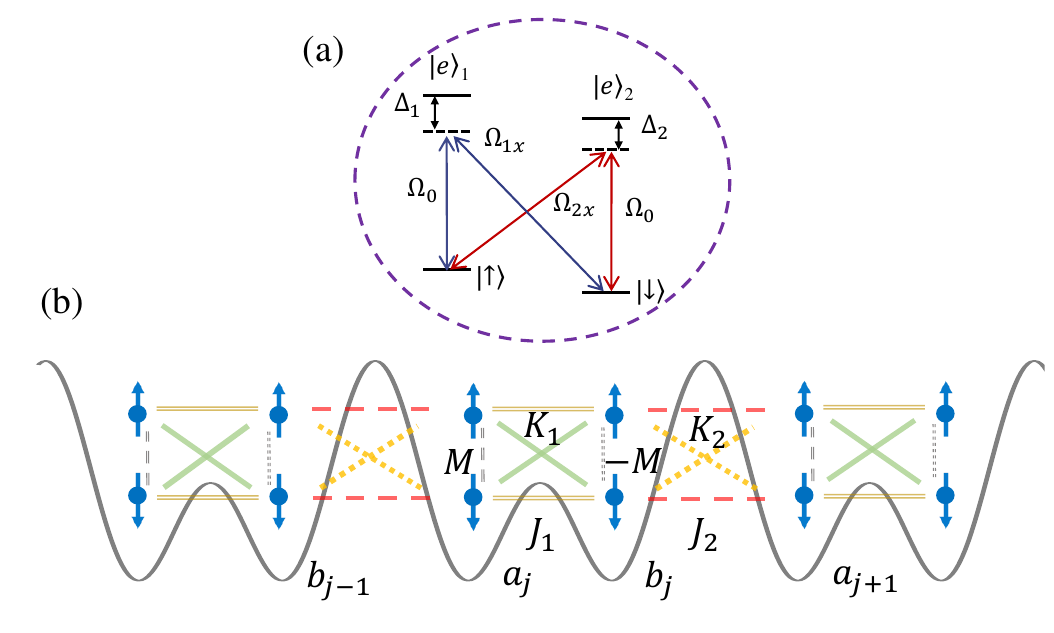}
	\caption{(a) Schematic illustration of the double-$\Lambda$ configuration for creating a Raman superlattice potential.
    Each Raman coupling is induced by one standing-wave laser with Rabi frequency $\Omega_{1x}$ or $\Omega_{2x}$ and one plane-wave laser with Rabi frequency $\Omega_0$. $\Delta_{1,2}$ are the detunings from the auxiliary excited states $|e\rangle_{1,2}$. (b) Implemented lattice model with dimerized spin-orbit couplings and staggered on-site spin flips. Specifically, each unit cell has two sublattice sites $a$ and $b$; the intra- and inter-cell spin-conserved hoppings are $J_{1,2}$, the intra- and inter-cell spin-flip hoppings are $K_{1,2}$ and the on-site spin-flip strengths for the sublattices $a$ and $b$ are $\pm M$. }
	\label{Fig1}
\end{figure}

\section{Optical Raman superlattices}

We consider two-component ultracold fermionic $^{173}$Yb (or $^{40}$K) atoms trapped in a one-dimensional optical Raman superlattice. The two components are represented by the two magnetic sublevels $\mid\uparrow\rangle=|5/2,3/2\rangle$ and $\mid\downarrow\rangle=|5/2,5/2\rangle$, respectively encoding the spin up and spin down. The optical Raman superlattice is produced by a spin-independent optical superlattice potential and a spin-dependent Raman superlattice potential. The state-independent optical superlattice potential is generated by superimposing two standing-wave lasers~\cite{SSHexp}, i.e,
$V_x=V_1\cos^2(k_0x)+V_2\cos^2(k_0x/2+\varphi/2)$, where $V_{1,2}$ and $\varphi$ can be varied by tuning the intensities and phases of the two standing-wave lasers. In contrast to previously applied Raman potentials~\cite{Liu2013,Liu2014,Pan2016,Liu2018,Pan2018,Song2018,CI2018,CI2019,Song2019,Pan2021}, here we consider a superlattice Raman potential created by two sets of two-photon Raman transitions. As illustrated in Fig. \ref{Fig1}(a), this is done by applying one standing-wave laser with Rabi frequency $\Omega_{1x}=\overline{\Omega}_{1}\sin(2k_0x)$ ($\Omega_{2x}=\overline{\Omega}_{2}\cos(k_0x+\theta)$) and one plane-wave laser with Rabi frequency $\Omega_0$ to induce the transition between the spin up (spin down) state and the auxiliary excited state $|e\rangle_1$ ($|e\rangle_2$). Note that such double-$\Lambda$ Raman transitions have been experimentally realized with $^{173}$Yb atoms~\cite{Song2018,Song2019}. When both transitions have a large detuning $\Delta_{1,2}$, we obtain two Raman potentials and their superposition creates a Raman superlattice potential $V_{so}\sigma_x$, where $V_{so}=\Omega_{1}\sin(2k_0x)+\Omega_{2}\cos(k_0x+\theta)$, $\Omega_{1,2}=\overline{\Omega}_{1,2}\Omega_0/\Delta_{1,2}$ and $\theta$ can be controlled by tuning the intensities and phases of the two Raman lasers.

The single-particle Hamiltonian for the optical Raman superlattice system is written as
\begin{equation}
H_{s}=\frac{p_x^2}{2m}+V_{x}+V_{so}\sigma_{x}.
\label{Hs}
\end{equation}
In the second quantization, this Hamiltonian takes the following form
\begin{equation}
H=\int dx\psi^{\dag}(x)H_s\psi(x)
\label{Hsq}
\end{equation}
where the field operator $\psi(x)=(\psi_{\uparrow}(x),\psi_{\downarrow}(x))^T$. Here we only consider atoms staying in the ground band. Then the field operator can be further expanded as
\begin{equation}
\psi_{\sigma}(x)=\sum_{x}W(x-j)C_{{j}\sigma},
\label{psi}
\end{equation}
where $C_{j\sigma}$ is the annihilation operator with spin $\sigma=\uparrow,\downarrow$ at the lattice site $j$ and $W(x-j)$ is the ground-band spin-independent Wannier function centered at the lattice site $j$. Here we assume the lattice spacing $a=\pi/k_0=1$. The tight-binding Hamiltonian for the optical Raman superlattice system is derived by substituting Eq. (\ref{psi}) into Eq. (\ref{Hsq})~\cite{Liu2013,Liu2014}.

In the case without Raman superlattice potential, the tight-binding Hamiltonian reads
\begin{align}
H_1=&-\sum_{j}t_{j,j+1}(C^{\dag}_{j\uparrow}C_{j+1\uparrow}+
C^{\dag}_{j\downarrow}C_{j+1\downarrow}+\text{H.c.}) \nonumber \\
&-\sum_{j}\delta_{j}(C^{\dag}_{j\uparrow}C_{j\uparrow}+C^{\dag}_{j\downarrow}C_{j\downarrow}).
\label{H_1}
\end{align}
The nearest-neighbor hopping rates are calculated as
\begin{align}
t_{j,j+1}=&-\int dx W^*(x-j)(\frac{p^2_x}{2m}+V_x)W(x-j-1)  \nonumber \\
=&t_0-V_2\int dxW^*(x-j)\dfrac{1+\cos(k_0x+\varphi)}{2}W(x-j-1))\nonumber\\
=&t_0+\frac{(-1)^j}{2}V_2\int dxW^*(x)\sin(k_0x+\varphi)W(x-1)\nonumber\\
=&t_0+\frac{(-1)^j}{2}V_2[\cos(\varphi)\int dxW^*(x)\sin(k_0x)W(x-1)\nonumber\\
 &+\sin(\varphi)\int dxW^*(x)\cos(k_0x)W(x-1)]\nonumber\\
=&t_0+(-1)^j(t_1\cos(\varphi)+t_2\sin(\varphi)),
\end{align}
where
\begin{align}
&t_0=-\int dx W^*(x-j)(\frac{p^2_x}{2m}+V_1\cos^2{(k_0x)})W(x-j-1),\nonumber \\
&t_1=\frac{V_2}{2}\int dx W^*(x)\sin(k_0x)W(x-1),\nonumber \\
&t_2=\frac{V_2}{2}\int dx W^*(x)\cos(k_0x)W(x-1).\nonumber \\
\label{t12}
\end{align}
The on-site energies are derived as
\begin{align}
\delta_{j}=&-\int dxW^*(x-j)(\frac{p^2_x}{2m}+V_x)W(x-j)  \nonumber \\
=&-V_2\int dx W^*(x-j) \dfrac{1+\cos(k_0x+\varphi)}{2}W(x-j)+\text{c.e.}  \nonumber \\
=&\frac{(-1)^j}{2}V_2\sin(\varphi)\int dx W^*(x)\cos(k_0x)W(x)+\text{c.e.} \nonumber \\
=&(-1)^{j}\delta\sin(\varphi)+\text{c.e.},
\end{align}
where
\begin{align}
\delta=\frac{V_2}{2}\int dx W^*(x)\cos(k_0x)W(x),
\label{t12}
\end{align}
$\text{c.e.}$ denotes constant energy and can be neglected. $\cos^2(k_0j)=0$ is used in the analytical calculations. As we can see, the parameters $t_{j,j+1}$ and $\Delta_j$ dependent on the parity of the lattice site $j$. When $\varphi=0,\pi$, the on-site energy vanishes, the corresponding system implements the SSH model~\cite{SSHexp}, but with spins, where nontrivial (trivial) topological phases could be prepared by tuning $\varphi=0$ ($\pi$).

Similarly, the tight-binding Hamiltonian created by the Raman superlattice potential is derived as
\begin{align}
H_2=&-\sum_{j}t^{so}_{j,j+1}(C^{\dag}_{j\uparrow}C_{j+1\downarrow}
+C^{\dag}_{j\downarrow}C_{j+1\uparrow}+\text{H.c.})\nonumber \\
&-\sum_{j}m_j(C^{\dag}_{j\uparrow}C_{j\downarrow}+\text{H.c.})
\label{H_so}
\end{align}
where the spin-orbit coupling strengths and on-site spin-flip rates are given by
\begin{align}
&t^{so}_{j,j+1}=t^{so}_0+(-1)^j(t^{so}_1\cos(\theta)+t^{so}_2\sin(\theta)), \nonumber \\
&m_j=(-1)^jm\sin(\theta)
\label{tso}
\end{align}
with
\begin{align}
&t^{so}_{0}=\Omega_1\int dx W^*(x)\sin(2k_0x)W(x-1), \nonumber \\
&t^{so}_1=\Omega_2\int dx W^*(x)\sin(k_0x)W(x-1),\nonumber \\
&t^{so}_2=\Omega_2\int dx W^*(x)\cos(k_0x)W(x-1),\nonumber \\
&m=\Omega_2\int dx W^*(x)\cos(k_0x)W(x).
\label{tso12}
\end{align}
We find that $t^{so}_{j,j+1}$ and $M_j$ also dependent on the parity of the lattice site $j$. When $\theta=0,\pi$, the Raman superlattice potential only generates dimerized spin-orbit couplings; When $\theta\neq0,\pi$, in addition to the dimerized spin-orbit couplings, staggered on-site spin flips are also induced by the Raman superlattice potential.

Due to the parity-dependent feature of the lattice parameters in both $H_1$ and $H_2$, each unit cell in the optical Raman superlattice has two sites. Suppose the two lattice sites in the $j$-th unit cell are labelled as $a_j$ and $b_j$. The total Hamiltonian $H=H_1+H_2$ can be rewritten as
\begin{align}
H=&\sum_{j}J_1(a^\dagger_{j\uparrow}b_{j\uparrow}+a^\dagger_{j\downarrow}b_{j\downarrow}+\text{H.c.})\nonumber\\
&+\sum_{j}J_2(a^\dagger_{j\uparrow}b_{j-1\uparrow}+a^\dagger_{j\downarrow}b_{j-1\downarrow}+\text{H.c.})   \nonumber\\
&+\Delta\sum_{j}(a^\dagger_{j\uparrow}a_{j\uparrow}+a^\dagger_{j\downarrow}a_{j\downarrow}
-b^\dagger_{j\uparrow}b_{j\uparrow}-b^\dagger_{j\downarrow}b_{j\downarrow})
\nonumber\\
&+\sum_{j}K_1(a^\dagger_{j\uparrow}b_{j\downarrow}+a^\dagger_{j\downarrow}b_{j\uparrow}+\text{H.c.})\nonumber\\
&+\sum_{j}K_2(a^\dagger_{j\uparrow}b_{j-1\downarrow}+a^\dagger_{j\downarrow}b_{j-1\uparrow}+\text{H.c.}) \nonumber\\
&+M\sum_{j}(a^{\dag}_{j\uparrow}a_{j\downarrow}-b^{\dag}_{j\uparrow}b_{j\downarrow}
+\text{H.c.}),
\label{H}
\end{align}
where $J_{1,2}=\pm(t_1\cos(\varphi)+t_2\sin(\varphi))-t_0$, $K_{1,2}=\pm(t^{so}_1\cos(\theta)+t^{so}_2\sin(\theta))-t^{so}_0$, $\Delta=\delta\sin(\varphi)$ and $M=m\sin(\theta)$. This model is highly tunable, in which the dimerized spin-conserved hoppings, dimerized spin-orbit couplings, staggered on-site spin filps and staggered on-site energies all can be individually controlled by tuning the laser intensities and phases, that allows the system into different topological phases.

\begin{figure*}
\centering
\includegraphics[width=16cm,height=8.5cm]{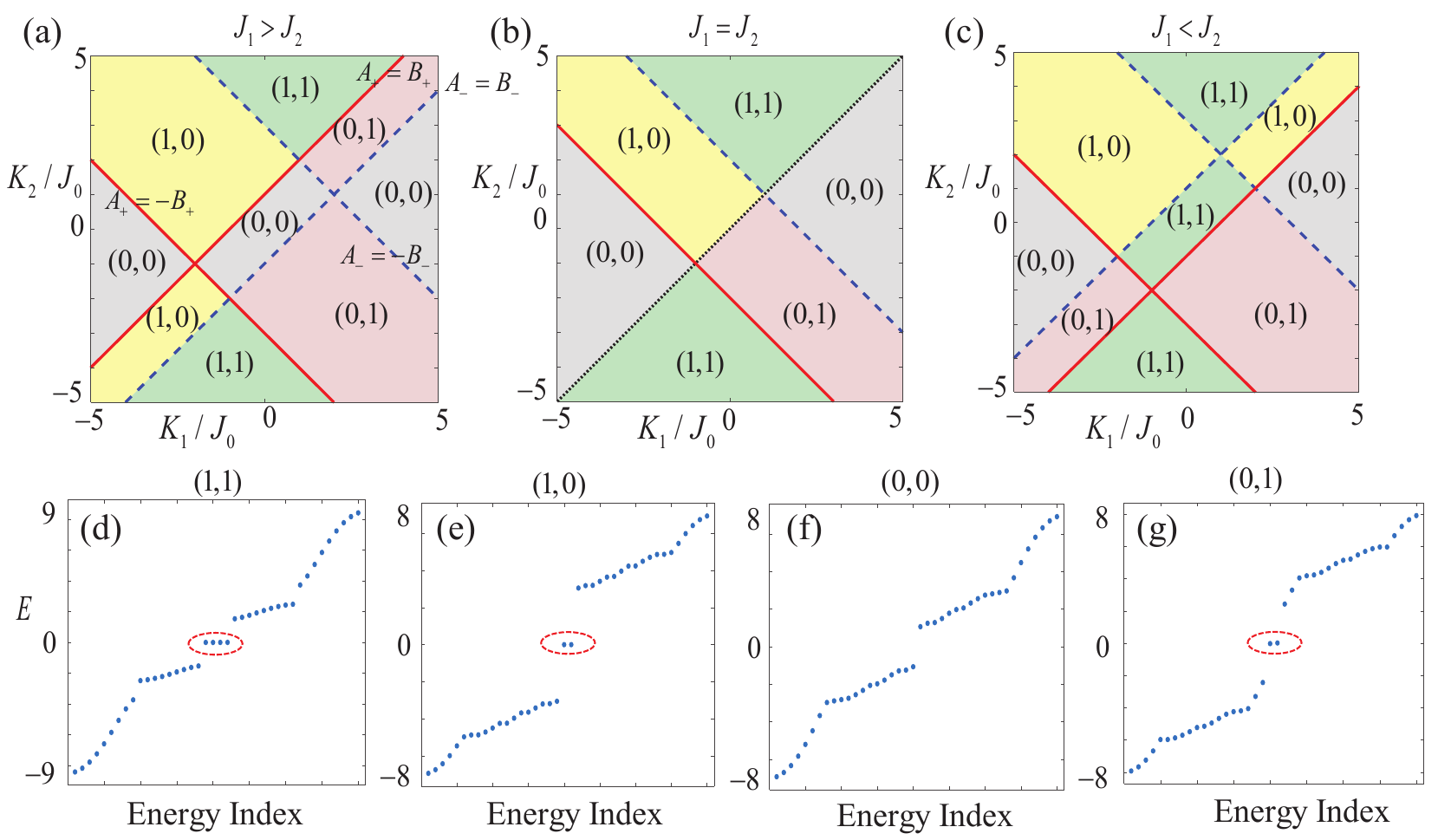}
\caption{Topological phase diagram and spin winding number values in the parameter space of $K_1/J_0$ and $K_2/J_0$ for (a) $J_1=2J_0$ and $J_2=J_0$, (b) $J_1=J_2=J_0$, (c) $J_2=2J_0$ and $J_1=J_0$. The transitions between different topological phases are determined by the gap closing conditions $E_{+}=0$ (red solid line, $A_{+}=\pm B_{+}$) and $E_{-}=0$ (blue dashed line, $A_{-}=\pm B_{-}$). The energy spectra with open boundary condition are shown in (d) $K_1=1.5J_0$ and $K_2=4J_0$, (e) $K_1=-3J_0$ and $K_2=3J_0$, (f) $K_1=3J_0$ and $K_2=1J_0$ and (g) $K_1=4J_0$ and $K_2=-3J_0$, corresponding to the four topological phases shown in (c), manifesting the bulk-edge correspondence. Here $J_0$ is used as energy unit.}
\label{Fig2}
\end{figure*}

\section{Spin-rotation symmetry}

Topological features of optical Raman superlattices are rooted in the momentum space. Through defining a four-component operator $C(k_x)=[a_{k_x\uparrow},a_{k_x\downarrow},b_{k_x\uparrow},b_{k_x\downarrow}]^T$ and implementing a Fourier transformation, the momentum-space Hamiltonian is derived as
$H(k_x)=\sum_{k_x}C^{\dag}(k_x)h(k_x)C(k_x)$, with
\begin{align}
h(k_x)&=(J_1+J_2\cos(k_x))\tau_x\sigma_{0}+J_2\sin(k_x)\tau_y\sigma_{0}  \nonumber\\
&+(K_1+K_2\cos(k_x))\tau_x\sigma_{x}+K_2\sin(k_x)\tau_y\sigma_{x}  \nonumber\\
&+\Delta\tau_z\sigma_{0}+M\tau_z\sigma_{x},
\label{hk}
\end{align}
where $\tau_i$ and $\sigma_i$ are the Pauli matrixes respectively defined on the sublattice and spin degree of freedom. We find that the Bloch Hamiltonian $h(k_x)$ satisfies a spin-rotation symmetry, i.e.,
\begin{align}
R_xh(k_x)R_x^{-1}=h(k_x),
\end{align}
where $R_x=\tau_0\sigma_x$ is the spin-rotation symmetry operator. As a result, the optical Raman lattice model is invariant under the spin rotation $(\uparrow,\downarrow)\rightarrow (\downarrow,\uparrow)$.

Since $[R_x,h(k_x)]=0$, the Bloch Hamiltonian $h(k_x)$ can be block-diagonalized in the eigenspace of $R_x$. i.e.,
\begin{align}
	\left|+ \right\rangle_1 =\dfrac{1}{\sqrt{2}}\left( {\begin{array}{*{20}{c}}
		1\\
		1\\
		0\\
		0
\end{array}} \right),	
\left|+ \right\rangle_2 =\dfrac{1}{\sqrt{2}}\left( {\begin{array}{*{20}{c}}
0\\
0\\
1\\
1
\end{array}} \right), \nonumber \\
\left|- \right\rangle_1 =\dfrac{1}{\sqrt{2}}\left( {\begin{array}{*{20}{c}}
		1\\
		-1\\
		0\\
		0
\end{array}} \right),
\left|- \right\rangle_2 =\dfrac{1}{\sqrt{2}}\left( {\begin{array}{*{20}{c}}
		0\\
		0\\
		1\\
		-1
\end{array}} \right),
\end{align}
with eigenvalues $\pm1$ respectively. In the following, we name the eigenvector spaces $\{|\pm\rangle_1,|\pm\rangle_2\}$ as the $\pm1$ spin-rotation subspaces. In such two subspaces, the Bloch Hamiltonian $h(k_x)$ is block-diagonalized into
\begin{align}
	\bar{h}(k_x)&=
  \left( {\begin{array}{*{20}{c}}
		h_+(k_x)&0\\
		0&h_-(k_x)\\
\end{array}} \right)
\label{dhk}	
\end{align}
where $\pm$ denote the $\pm1$ spin-rotation subspaces. The block Hamiltonian takes the following form
\begin{align}
h_{\pm}(k_x)=&d_{x\pm}s_{x\pm}+d_{y\pm}s_{y\pm}+d_{z\pm}s_{z\pm},
\label{blockH}
\end{align}
with $d_{x\pm}=A_{\pm}+B_{\pm}\cos(k_x)$, $d_{y\pm}=B_{\pm}\sin(k_x)$ and $d_{z\pm}=\Delta\pm M$, where $A_{\pm}=J_1\pm K_1$, $B_{\pm}=J_2\pm K_2$ and $s_{x\pm,y\pm,z\pm}$ are the Pauli matrixes defined in the $\pm1$ spin-rotation subspaces. The eigenvalues of the block Hamiltonian $h_{\pm}(k_x)$ are
\begin{equation}
E_{s}=\pm\sqrt{A^2_{s}+B^2_{s}+(\Delta+sM)^2
+2A_{s}B_{s}\cos(k_x)},
\label{Ek}
\end{equation}
where $s=\pm$.

Below we will demonstrate that the presence of the spin-rotation symmetry simplifies the characterization of the topology of $h(k_x)$. We exhibit that the four-band topological features associated with $h(k_x)$ can be characterized through two spin-dependent topological invariants, that are defined in the $\pm1$ spin-rotation subspaces based on $h_{\pm}(k_x)$.

\begin{figure*}
\centering
\includegraphics[width=17cm,height=7cm]{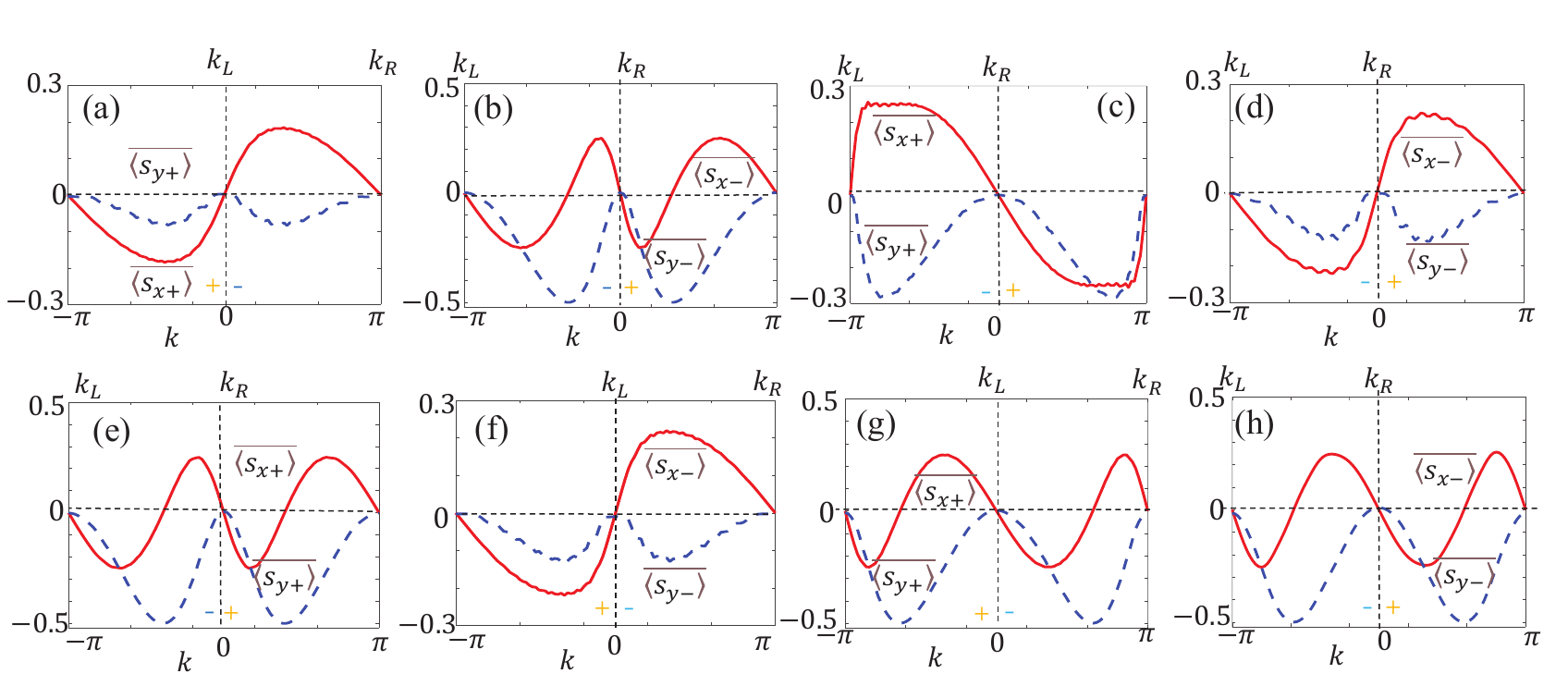}
\caption{Time-averaged spin polarizations $\overline{\langle s_{x\pm,y\pm}\rangle}$ as a function of $k_x$ for (a,b) $K_1=4J_0$ and $K_2=-4J_0$; (c,d) $K_1=3J_0$ and $K_2=J_0$; (e,f) $K_1=-3J_0$ and $K_2=4J_0$; (g,h) $K_1=1.5J_0$ and $K_2=4J_0$. The sign $+$ ($-$) denotes the region where $d_{y\pm}>0$ ($d_{y\pm}<0$). The other parameters are $J_1=J_0$, $J_2=2J_0$ and $T=10/J_0$. }
\label{Fig3}
\end{figure*}

\section{Topological phases protected by spin winding numbers}

We firstly study the case without on-site energies and on-site spin flips by tuning the laser phases $\varphi=\theta=0$. In this case, the block Hamiltonian $h_{\pm}(k_x)$ respect a chiral symmetry, allowing us to employ winding numbers to characterize their topology. In each spin-rotation subspace, a spin winding number can be defined, i.e.,
\begin{align}
\nu_{\pm} =\frac{1}{2\pi}\int dk_{x}\mathbf{n_{\pm}}\times \partial _{k_{x}}\mathbf{n_{\pm}},
\label{swn}
\end{align}
where $\mathbf{n_{\pm}}=(d_{x\pm},d_{y\pm})/(d^2_{x\pm}+d^2_{y\pm})^{1/2}$. The spin winding numbers $\nu_{+}$ and $\nu_{-}$ separately characterize the topology of the block Hamiltonian $h_{+}(k_x)$ and $h_{-}(k_x)$. By substituting Eq. (\ref{blockH}) into Eq. (\ref{swn}), the spin winding numbers are calculated as
\begin{equation}
    \nu_{\pm}=
   \begin{cases}
   1 &\mbox{if $|A_{\pm}|<|B_{\pm}|$}\\
   0 &\mbox{otherwise}.
   \end{cases}
   \label{swn2}
  \end{equation}

The topological phase diagrams in terms of ($\nu_{+},\nu_{-}$) are studied in Figs. \ref{Fig2}(a-c). We can find that the Raman superlattice system features four different chiral topological phases. The transitions between different topological phases, signified by the change of spin winding number values, are usually accompanied by gap closings. Through examining the gap closings in Figs. \ref{Fig2}(a-c), we find that the optical Raman lattice supports abundant topological phase transitions. For the bloch Hamiltonian $h_{\pm}(k_x)$, the gap closing conditions are $E_{\pm}=0$ and give the gap closing lines $|A_{\pm}|=|B_{\pm}|$. The topological invariant $\nu_{+}$ ($\nu_{-}$) would change once crossing the gap closing lines $E_{+}=0$ ($E_{-}=0$), agreeing well with the theoretical predictions by Eq. (\ref{swn2}).

Fig. \ref{Fig2}(a) investigates the spin winding numbers as a function of the spin-orbit coupling strengthes for $J_1>J_2$. When the spin-orbit couplings are turned off, the corresponding optical Raman lattice is described by two independent SSH models and in the trivial topological phases, with the topological invariants as ($\nu_{+}=0,\nu_{-}=0$). When the spin-orbit couplings are turned on, with the increase of $K_{1,2}$, the energy gap would firstly close at $E_{+}=0$ ($A_{+}=\pm B_{+}$) or $E_{-}=0$ ($A_{-}=\pm B_{-}$); After the gap reopening, the system will come into a nontrivial topological phase with ($\nu_{+}=1,\nu_{-}=0$) or ($\nu_{+}=0,\nu_{-}=1$). While if the energy gap successively closes at $E_{+}=0$ and $E_{-}=0$, the final topological phase is ($\nu_{+}=1,\nu_{-}=1$).

Fig. \ref{Fig2}(b) shows that the gap closing lines $A_{+}=B_{+}$ and $A_{-}=B_{-}$ coincide for $J_1=J_2$. In contrast to Fig. \ref{Fig2}(a), the region between such two lines vanishes, and topological phases transitions could take place directly between ($\nu_{+}=1(0),\nu_{-}=1$) and ($\nu_{+}=0(1),\nu_{-}=0$). It is also worth pointing out that in this case the topological features mainly comes from the Raman superlattice. Fig. \ref{Fig2}(c) further displays that, for $J_1<J_2$, the region between the gap closing lines $A_{+}=B_{+}$ and $A_{-}=B_{-}$ reopens, in which the three phases simultaneously undergo a topological phase transition as compared to Fig. \ref{Fig2}(a). As a consequence, when the spin-orbit coupling is fixed, tuning $J_{1,2}$ could also drive the system into different topological phases.

Figs. \ref{Fig2}(d-g) exhibit the bulk-edge correspondence associated with the four topological phases ($\nu_{+}=0,1,\nu_{-}=0,1$) shown in Fig. \ref{Fig2}(c). The edge modes guaranteed by the bulk topological invariants are investigated by calculating the energy spectra of $H$, with open boundary condition. For the topological phase with ($\nu_{+}=1,\nu_{-}=1$), Fig. \ref{Fig2}(d) shows that there are four degenerate zero-energy topological edge states, i.e., two left and two right zero-energy edge states. The wave functions (not normalized) for the four edge states are directly calculated as~\cite{TIBook,Mei2018}
\begin{align}
|\psi^{L}_{\pm}\rangle =\sum_{j=1}^{N}(-1)^{j}
\left( \frac{A_{\pm}}{B_{\pm}}\right) ^{j}
\frac{a_{j\uparrow}^{+}\pm a_{j\downarrow}^{+}}{\sqrt{2}}|0\rangle, \nonumber \\
|\psi^{R}_{\pm}\rangle =\sum_{j=1}^{N}(-1)^{j-N}
\left( \frac{A_{\pm}}{B_{\pm}}\right)^{j-N}
\frac{b_{j\uparrow}^{+}\pm b_{j\downarrow}^{+}}{\sqrt{2}}|0\rangle.
\end{align}%
While for ($\nu_{+}=1(0),\nu_{-}=0(1)$), there are only two zero-energy topological edge states (see Figs. \ref{Fig2}(e,g)), i.e., one left edge state $|\psi^{L}_{+}\rangle$ ($|\psi^{L}_{-}\rangle$) and one right edge states $|\psi^{R}_{+}\rangle$ ($|\psi^{R}_{-}\rangle$). By contrast, for ($\nu_{+}=0,\nu_{-}=0$), there are no zero-energy edge states (see Fig. \ref{Fig2}(f)).

The winding numbers could be measured through quench dynamics~\cite{Zhang2018,Zhang2019a}. Here we show that a single quench process could be used to simultaneously measure the two spin winding numbers $\nu_{\pm}$. Suppose the initial state of the system is prepared into an equal superposition of the ground state of the block Hamiltonian $h_{\pm}=d_{x\pm}s_{x\pm}+(d_{y\pm}+m_{y\pm})s_{y\pm}$ with $m_{y\pm}\gg1$, i.e., $|\psi(t=0)\rangle=(|+\rangle_1-i|+\rangle_2+|-\rangle_1-i|-\rangle_2)/2$. After that, the Hamiltonian governing the time evolution is $h(k_x)$ that is equivalent to $h_{\pm}(k_x)$ with $m_{y\pm}=0$, which thus accomplishes a $y$-direction quantum quench for both block Hamiltonian, i.e., from $m_{y\pm}\gg1$ to $m_{y\pm}=0$. From the measured time-resolved spin polarizations $\langle s_{x\pm,y\pm}(k_x)\rangle_{t}=\langle\psi(t=0)|e^{ih(k_x)t}s_{x\pm,y\pm} e^{-ih(k_x)t}|\psi(t=0)\rangle$, the time-averaged spin polarizations are extracted as $\overline{\langle s_{x\pm,y\pm}(k_x)\rangle}=\frac{1}{T}\int_{0}^{T}dt\langle s_{x\pm,y\pm}(k_x)\rangle_t$. When $T$ is longer enough, the spin winding numbers can be measured through
\begin{align}
	v_{\pm}=\dfrac{1}{2}(g_{x\pm}(k_R)-g_{x\pm}(k_L))
	\label{swnm}
\end{align}
where $g_{x\pm}(k_{L,R})=-\text{sgn}(\partial_{k_\perp}\overline{\langle s_{x\pm}\rangle})$~\cite{Zhang2018,Zhang2019a}, with $k_{L,R}$ as the band inversion surfaces (BISs) given by $\overline{\langle s_{y\pm}(k_{L,R})\rangle}=0$, and $k_\perp$ as the momentum pointing from the region $d_{y\pm}<0$ to $d_{y\pm}>0$.

Fig. \ref{Fig3} presents the time-averaged spin polarizations as a function of $k_x$ for different lattice parameters. According to Fig. \ref{Fig2}(c), the spin winding numbers corresponding to the lattice parameters in Figs. \ref{Fig3}(a,b) are ($\nu_{+}=0,\nu_{-}=1$). Figs. \ref{Fig3}(a,b) clearly show that, the time-averaged spin polarizations $\overline{\langle s_{y\pm}\rangle}$ vanish at the BISs, yielding the locations of BISs as $k_x=k_{L,R}$; The time-averaged spin polarizations near the BISs give $g_{x+}(k_{R})=g_{x+}(k_{L})=-1$ and $g_{x-}(k_{R})=-g_{x-}(k_{L})=1$. Based on Eq. (\ref{swnm}), the spin winding numbers are measured as $\nu_{+}=0$ and $\nu_{-}=1$, fully agreeing with theoretical predicted values. The results in Figs. \ref{Fig3}(c,d) show that $g_{x+}(k_{R})=g_{x+}(k_{L})=1$ and $g_{x-}(k_{R})=g_{x-}(k_{L})=1$, giving the the spin winding numbers $\nu_{+}=0$ and $\nu_{-}=0$. Similarly, the spin winding numbers measured in Figs. \ref{Fig3}(e,f) and (g,h) are respectively ($\nu_{+}=1,\nu_{-}=0$) and ($\nu_{+}=1,\nu_{-}=1$).

\section{Topological phases protected by spin Chern numbers}

We further find that, by associating the laser phases $\varphi=\theta$ ($\in(0,2\pi)$) with a synthetic momentum, the optical Raman superlattice is described by $h(k_x,\varphi)$ and provides a natural platform for exploring two-dimensional quantum spin Hall insulator phases, where the two-dimensional Brillouin zone is defined by the genuine momentum $k_x\in(0,\pi]$ and the synthetic momentum $\varphi\in(0,2\pi]$. In this case, the block Hamiltonian $h_{\pm}(k_x)$ are mapped into $h_{\pm}(k_x,\varphi)$, and the corresponding topology are characterized by spin Chern numbers instead, defined as
\begin{equation}
C_{\pm}=\frac{1}{4\pi }\int \int dk_{x}d\varphi (\partial_{k_{x}}\mathbf{n_{\pm}}\times
\partial_{\varphi}\mathbf{n_{\pm}})\cdot \mathbf{n_{\pm}},
\end{equation}
where $\mathbf{n_{\pm}}=(d_{x\pm},d_{y\pm},d_{z\pm})/(d^2_{x\pm}+d^2_{y\pm}+d^2_{z\pm})^{1/2}$. As indicated in Eqs. (\ref{t12},\ref{tso12}), the ratios of the lattice parameters $\alpha=t^{so}_0/t_0$ and $\beta=t^{so}_1/t_1=t^{so}_2/t_2=m/\delta$ can be flexibly controlled through tuning $V_{1,2}$ and $\Omega_{1,2}$. Below we will show that tuning $\alpha$ and $\beta$ allows us to explore different two-dimensional topological phases.

\begin{figure*}
\centering
\includegraphics[width=16cm,height=7.5cm]{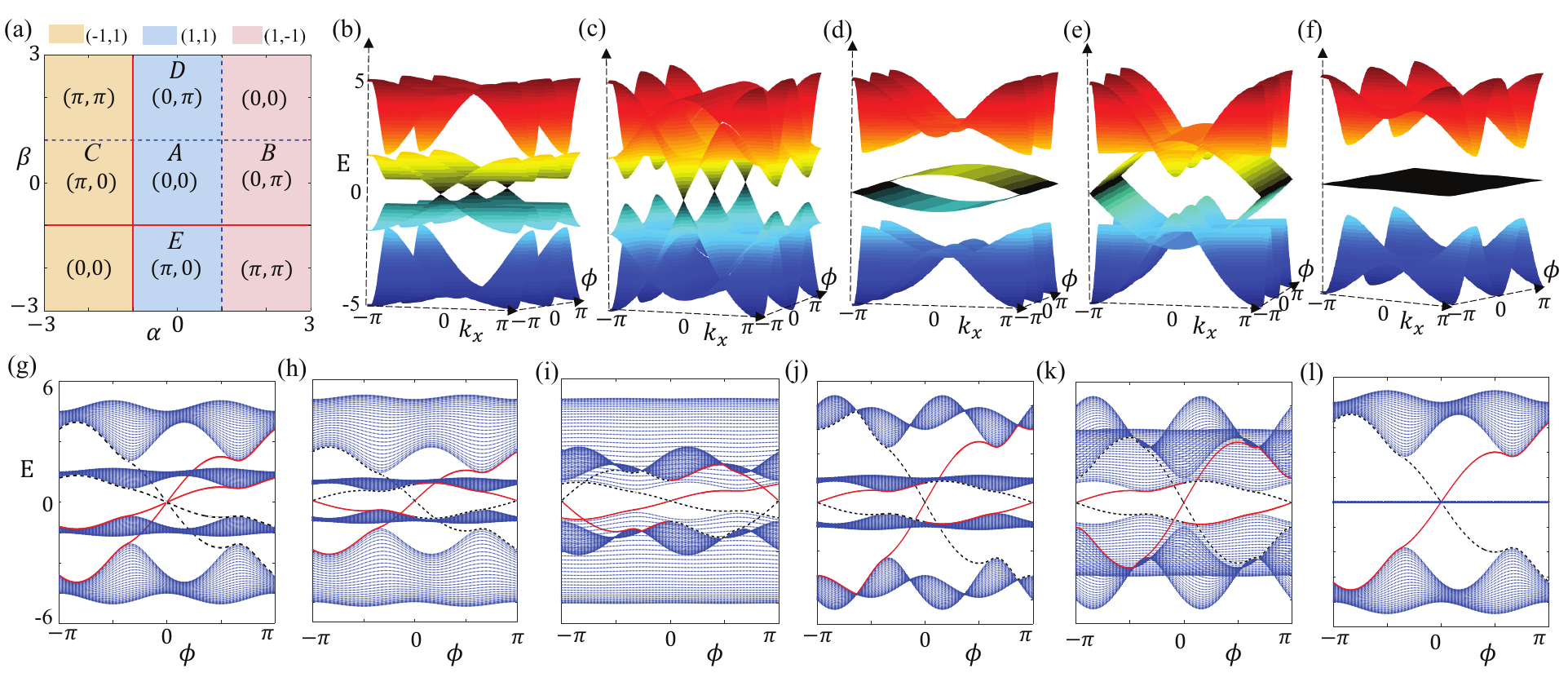}
\caption{(a) Topological phase diagram and spin Chern number values in the parameter space of $\alpha$ and $\beta$. The solid (dashed) lines $\alpha=-1$ and $\beta=-1$ ($\alpha=1$ and $\beta=1$) are the gap closing lines in which $E_{+}=0$ ($E_{-}=0$). The integer values in the coordinates represent the spin Chern number values $C_{\pm}$, while the values of the laser phase $\varphi$ in the coordinates reflect the positions at which the edge states of $h_{\pm}$ cross. For example, $(1,-1)$ represents that the spin Chern number values are $(C_{+}=1,C_{-}=-1)$; $(0,\pi)$ represents that the edge states of $h_{+}$ and $h_{-}$ in this region of parameter space respectively cross at $\varphi=0$ and $\pi$ (see (g-k)). The bulk energy spectra in the synthetic first Brillouin zone for (b) $\alpha=1$ and $\beta=0.5$, (c) $\alpha=-1$ and $\beta=0.5$, (d) $\alpha=0.5$ and $\beta=-1$, (e) $\alpha=0.5$ and $\beta=-1$ and (f) $\alpha=1$ and $\beta=1$. With open boundary condition along the genuine dimension, the corresponding energy spectra in the five regions of parameter space $A-E$ (labelled in (a)) are respectively plotted in (g-k) and for the flat band case is shown in (l). The specified parameters are (g) $\alpha=0.5$ and $\beta=0.5$, (h) $\alpha=1.5$ and $\beta=0.5$, (i) $\alpha=-1.5$ and $\beta=0.5$, (j) $\alpha=0.5$ and $\beta=1.5$, (k) $\alpha=0.5$ and $\beta=-1.5$ and (l) $\alpha=1$ and $\beta=1$. The red solid and black dashed lines respectively denote the left and right in-gap topological edge states. The other parameters are $t_1=0.8t_0$, $t_2=0.3t_0$ and $\delta=-t_0$. Here $t_0$ is used as energy unit.}
\label{Fig4}
\end{figure*}

According to the spin Chern numbers $(C_{+},C_{-})$, the topological phase diagram in the parameter space of $\alpha$ and $\beta$ is obtained in Fig. \ref{Fig4}(a). The values of $(C_{+},C_{-})$ are calculated as $C_{+}=1(-1)$ for $\alpha>-1$ ($\alpha<-1$) and $C_{-}=1(-1)$ for $\alpha>1$ ($\alpha<1$). From which we find that, there are two different quantum spin Hall insulator phases identified by $(C_{+}=-1,C_{-}=1)$ and $(C_{+}=1,C_{-}=-1)$, separated by a double Chern insulator phases identified by $(C_{+}=1,C_{-}=1)$. Through the gap closings, as plotted in Fig. \ref{Fig4}(a), we also find that the topological phase transitions in this synthetic two-dimensional system have an interesting characteristic. The gap closing conditions for $h_{\pm}$ are $E_{\pm}=0$, respectively giving the gap closing lines $\alpha=\mp1$ and $\beta=\mp1$. As we can see, topological phase transitions do occur at the gap closing lines $\alpha=\mp1$, signified by the change of the spin Chern number values $C_{\pm}$ crossing theses lines. However, there is no topological phase transitions at the gap closing lines $\beta=\mp1$, across which the spin Chern numbers $C_{\pm}$ are same.

As shown in Figs. \ref{Fig4}(b-e), this is attributed to that the physical mechanism for producing the gap closings is quite different. For $\alpha=\pm1$, the energy gaps close at the two-dimensional Dirac points (see Figs. \ref{Fig4}(b-c)), accompanied by band inversions after the reopening of the gaps, leading to topological phase transitions. While for $\beta=\pm1$, the energy spectra for the middle two bands have nothing to do with the synthetic momentum $\varphi$ (see Figs. \ref{Fig4}(d-e)), and the energy gap closes at one-dimensional Dirac points $k_x=\pm\pi$. This means that, the couplings along the synthetic dimension are turned off and the system is decoupled into independent one-dimensional lattices in the genuine dimension. Consequently, band inversions and topological phase transitions do not take place in the two-dimension momentum space in this situation. In addition, the energy bands for $\alpha=\beta=\pm1$ present an interesting feature, where the middle two energy bands merge into degenerate zero-energy flat bands, with the sum of spin Chern numbers as zero, as depicted in Fig. \ref{Fig4}(f).

Figs. \ref{Fig4}(g-k) plot the energy spectra of edge states corresponding to the five regions (labelled by A-E respectively) in the topological phase diagram. The results show that, there are one pair of in-gap edge states connecting the energy bands of $h_{+}$ and $h_{-}$ respectively, agreeing with the prediction by the bulk-edge correspondence and the corresponding spin Chern number values $C_{\pm}$. Moreover, the crossing points for the left (red solid lines) and right (black dotted lines) edge states of $h_{+}$ and $h_{-}$ both cross at $\varphi=0$ or $\pi$, which has been specifically given in Fig. \ref{Fig4}(a). For example, in the region B, $(0,\pi)$ denotes that the edge states of $h_{+}$ and $h_{-}$ respectively cross at $\varphi=0$ and $\pi$, as shown in Fig. \ref{Fig4}(h). We also notice that, for the edge states of $h_{+}$ ($h_{-}$), once crossing the gap closing lines $E_{+}=0$ ($E_{-}=0$), the crossing points would change from $\varphi=0(\pi)$ to $\varphi=\pi(0)$, regardless of whether topological phase transition occurs. The differences between different topological phases are manifested by the group velocities of the edge states. For instance, by making a comparison between Fig. \ref{Fig4}(g) and (h), we can observe that the group velocities for the left or right edge state of $h_{-}$ ($h_{+}$) are opposite (same), reflecting that the sign of the spin Chern numbers $C_{-}$ ($C_{+}$) are opposite (same), revealing the difference between the topological phases $(C_{+}=1,C_{-}=1)$ and $(C_{+}=1,C_{-}=-1)$. The edge states for $\alpha=\beta=\pm1$ are presented in Fig. \ref{Fig4}(l), where there is only one pair of edge states, due to the merging of the two middle bands into a single topologically trivial flat band.

\begin{figure*}
\centering
\includegraphics[width=14.3cm,height=6.5cm]{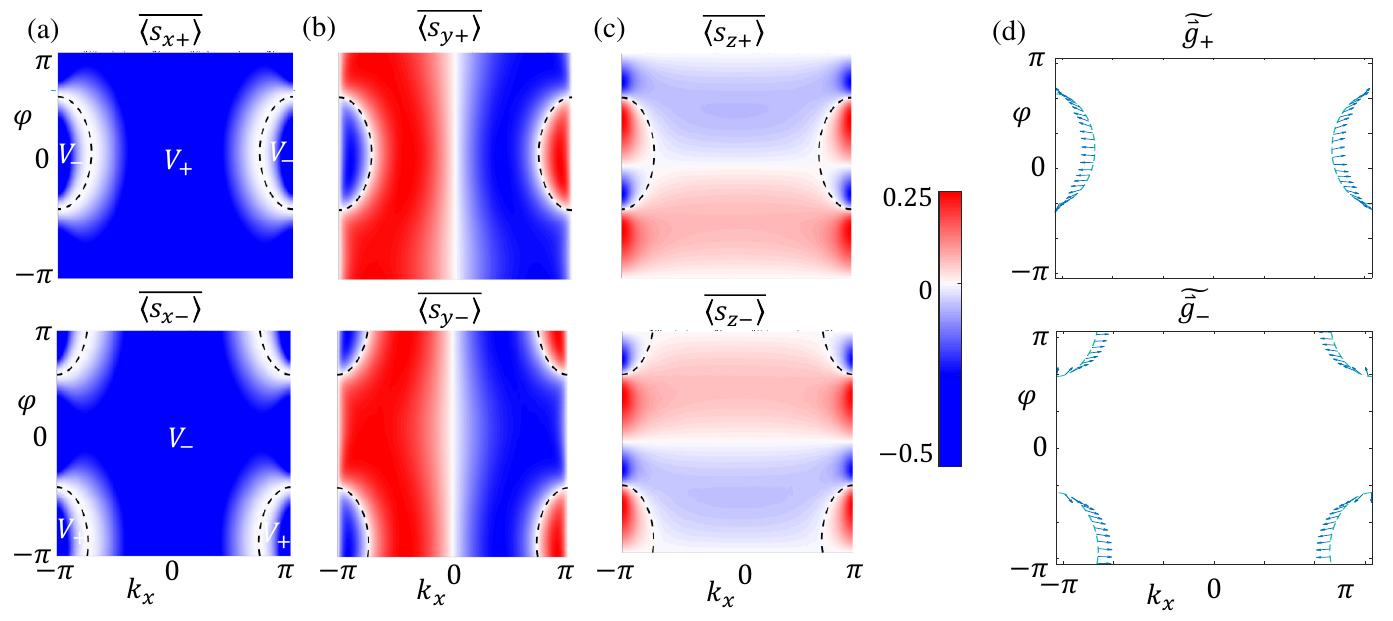}
\caption{(a-c) Time-averaged spin polarizations $\overline{\langle s_{x\pm,y\pm,z\pm}\rangle}$ as a function of $k_x$ and $\varphi$. The black dashed lines denote the BISs that divide the synthetic Brillouin zone into two regions, i,e., $\mathcal{V}_{+}$ with $d_{x\pm}>0$ and $\mathcal{V}_{-}$ with $d_{x\pm}<0$. The spin textures for the dynamic fields $\vec{g}_{\pm}$ on the BISs are given in (d). The parameters are chosen as $t_1=0.8t_0$, $t_2=0.3t_0$, $\delta=-t_0$, $\alpha=4$, $\beta=0.3$, and $T=10/t_0$.}
\label{Fig5}
\end{figure*}

Quench dynamics also allows to measure Chern numbers~\cite{Zhang2018,Zhang2019b}. We show that the two spin Chern numbers $C_{\pm}$ can be simultaneously measured from a single quench process. As an example, the measurement of $(C_{+}=1,C_{-}=-1)$ is exhibited as follows. Similar to measure spin winding numbers, here we perform a $x$-direction quantum quench by initially preparing the system into $|\psi(t=0)\rangle=(|+\rangle_1-|+ \rangle_2 +|-\rangle_1-|-\rangle_2 )/2$ and letting it evolve under $h(k_x,\varphi)$. After that, the time-averaged spin polarizations $\overline{\langle s_{x\pm,y\pm,z\pm}\rangle}$ are measured, as numerically shown in Figs. \ref{Fig4}(a-c). From them, the BISs for $h_{\pm}$ are determined by $d_{x\pm}=0$ and measured through $\overline{\langle s_{x\pm}\rangle}=0$; the dynamic fields $\vec{g}_{\pm}=(g_{y\pm},g_{z\pm})$ are extracted by $g_{i\pm}=-\partial_{k_\perp}\overline{\langle s_{i\pm}\rangle}$~\cite{Zhang2018,Zhang2019b}, with $k_\perp$ as the momentum pointing from the region $\mathcal{V}_{-}$ to $\mathcal{V}_{+}$. Fig. \ref{Fig5}(d) shows the dynamic fields $\vec{g}_{\pm}$ on the BISs. The Chern number is given by the winding number of the dynamic fields along the BISs~\cite{Zhang2018,Zhang2019b}. As plotted in Fig. \ref{Fig5}(d), the dynamic fields $\vec{g}_{\pm}$ both wind the BISs one time, but with opposite winding directions, giving the spin Chern numbers $C_{+}=1$ and $C_{-}=-1$ respectively.

\section{Summary and Outlook}

In summary, we have proposed an experimentally relevant optical Raman superlattice system and systematically studied its topological properties. We have found that such system respects a spin-rotation symmetry that allows us to characterize its topological property through spin winding and spin Chern numbers, which could be directly measured by nonequilibrium quench dynamics. We have further exhibited that this system features various topological phases, such as the four-band chiral topological insulator phases, quantum spin Hall insulator phases and double Chern insulator phases, with several interesting topological features, like the multifarious topological phase transitions, tunable zero-energy modes and degenerate zero-energy flat bands.

The results in our study clearly show that, due to the interplay between optical superlattice and Raman superlattice, optical Raman superlattice system has much richer topological properties, which could provide more opportunities for implementing and probing topological phases of ultracold atoms. For example, when the optical superlattice system is prepared into the quantum spin Hall insulator phases, $Z_2$ topological pumping~\cite{Kane2006} can be naturally implemented by adiabatically controlling the laser phase over one period. In the near future, it is quite interesting to generalize optical Raman superlattices to two and three dimensions for exploring diverse high-dimensional topological phases, including the topological semimetal phases~\cite{Young2012,Young2015,Zhao2016}, higher-order topological insulator phases~\cite{Hughes2017,Neupert2018} and four-dimensional topological insulator phases~\cite{Bloch2018,Zilberberg2013,Goldman2015}.

\section{Acknowledgment}

We thank H.Q. Wang and M.N. Chen for helpful discussions. This work was supported by the National Key Research and Development Program of China (2017YFA0304203), National Natural Science Foundation of China (NSFC) (12034012, 12074234), Changjiang Scholars and Innovative Research Team in University of Ministry of Education of China (PCSIRT)(IRT\_17R70), Fund for Shanxi 1331 Project Key Subjects Construction, and 111 Project (D18001).

\end{document}